\documentclass[a4paper]{jpconf}

\usepackage{graphicx}
\usepackage{amsmath,amssymb}
\usepackage{siunitx}
\usepackage{hyperref}
\usepackage{cite}
\newcommand{\unit}[1]{\ {\rm #1}}

\begin{document}
\title{First observation and analysis of DANCE: 
Dark matter Axion search with riNg Cavity Experiment}

\author{Yuka Oshima$^1$, Hiroki Fujimoto$^1$, Masaki Ando$^{1, 2}$, \\
Tomohiro Fujita$^{2, 3}$, Jun'ya Kume$^{1, 2}$, Yuta Michimura$^{1, 4}$, \\
Soichiro Morisaki$^5$, Koji Nagano$^6$,  Hiromasa Nakatsuka$^7$, \\
Atsushi Nishizawa$^2$, Ippei Obata$^8$, Taihei Watanabe$^1$}

\address{$^1$ Department of Physics, University of Tokyo, Bunkyo, Tokyo 113-0033, Japan}
\address{$^2$ Research Center for the Early Universe, University of Tokyo, Bunkyo, Tokyo 113-0033, Japan}
\address{$^3$ Waseda Institute for Advanced Study, Waseda University, Shinjuku, Tokyo 169-8050, Japan}
\address{$^4$ PRESTO, Japan Science and Technology Agency (JST), Kawaguchi, Saitama 332-0012, Japan}
\address{$^5$ Department of Physics, University of Wisconsin-Milwaukee, Milwaukee, Wisconsin 53201, USA}
\address{$^6$ Institute of Space and Astronautical Science, Japan Aerospace Exploration Agency, Sagamihara, Kanagawa 252-5210, Japan}
\address{$^7$ Institute for Cosmic Ray Research, University of Tokyo, Kashiwa, Chiba 277-8582, Japan}
\address{$^8$ Max-Planck-Institut f\"{u}r Astrophysik, Karl-Schwarzschild-Stra\ss e 1, 85741 Garching, Germany}

\ead{yuka.oshima@phys.s.u-tokyo.ac.jp}

\begin{abstract}
Dark matter Axion search with riNg Cavity Experiment (DANCE)
was proposed to search for axion dark matter
[{\em Phys. Rev. Lett.\/} \href{https://doi.org/10.1103/PhysRevLett.121.161301}{{\bf 121}, 161301 (2018)}].
We aim to detect the rotation and oscillation of optical linear polarization
caused by axion-photon coupling with a bow-tie cavity.
DANCE can improve the sensitivity 
to axion-photon coupling constant $g_{a \gamma}$ 
for axion mass $m_a < 10^{-10}~\si{eV}$ 
by several orders of magnitude
compared to the best upper limits at present.
A prototype experiment DANCE Act-1 is ongoing
to demonstrate the feasibility of the method and
to investigate technical noises.
The optics was assembled and the performance of the cavity was evaluated.
The first 12-day observation was successfully performed in May 2021.
We reached $3 \times 10^{-6} \unit{rad/\sqrt{Hz}}$ at $\SI{10}{Hz}$
in the one-sided amplitude spectral density of the rotation angle of linear polarization.
\end{abstract}

\section{Introduction}

Various experiments and observations have been performed to search for dark matter,
but dark matter has yet to be detected.
Axions are one of the well-motivated candidates for dark matter
since axions behave like non-relativistic classical wave fields
in the present universe~\cite{QCD1, QCD2, QCD3, ALP}.
Axions may weakly interact with photons~\cite{theory, theory2},
and this axion-photon coupling provides a good chance 
to detect axions through direct search experiments
by using well-developed photonics technology.
Recently, several novel methods were proposed
to observe axion-photon coupling
using carefully designed optical cavities~\cite{AxionInterferometry, DANCE, ADBC, KAGRA, aux, ADAMGD}.
These laser interferometric searches can be done
without a strong magnetic field,
and have good sensitivity in the low mass region $m_a < 10^{-10}~\si{eV}$.
In this paper,
we review the Dark matter Axion search with riNg Cavity Experiment (DANCE) proposal, and
report the status of the prototype experiment, DANCE Act-1.

\section{Designed sensitivity of DANCE}
The axion-photon interaction gives
a phase velocity difference between left- and
right-handed circularly polarized light~\cite{theory, theory2}.
The phase velocity difference 
$\delta c = |c_{\rm{L}} - c_{\rm{R}}| 
= \delta c_0 \sin(m_a t + \delta_{\tau} (t))$
with axion mass $m_a$ and a phase factor $\delta_{\tau} (t)$
for a wavelength of light
$\lambda = 2\pi /k$
is estimated to be
\begin{equation}
\delta c_0 = \frac{g_{a \gamma} a_0 m_a}{k} 
\simeq 2.1 \times 10^{-24} \left( \frac{\lambda}{\SI{1064}{nm}} \right)
\left( \frac{g_{a \gamma}}{10^{-12} \unit{GeV^{-1}}} \right).
\end{equation}
Here, we assumed axion energy density equals local dark matter density, 
$\rho_a = m_a^2 a_0^2 /2 \simeq 0.4 \unit{GeV/cm^3}$.

This phase difference between circular polarizations is
equivalent to a rotation of linearly polarized light~\cite{theory}.
Small signal sidebands are generated 
as a linearly polarized laser light propagates
in the presence of axions~\cite{ADBC}.
Optical path length can be effectively increased using an optical cavity
and the amplitude of the sidebands is enhanced for detection.
The polarization flip upon
mirror reflection have to be taken into account
when designing the optical cavities.
A bow-tie ring cavity is used 
to prevent the linear polarization from inverting
since the laser beam is reflected twice at both ends (see Figure \ref{fig:setup} (a))~\cite{DANCE}.

The fundamental noise source of DANCE would be quantum shot noise.
The one-sided amplitude spectral density of the shot noise is given by
\begin{equation}
\sqrt{S_{\rm{shot}} (\omega)} =
\sqrt{\frac{\hbar \lambda}{4\pi c P_{\rm{trans}}}
\left( \frac{1}{t_c^2} + \omega^2 \right)},
\label{eq:shotnoise}
\end{equation}
where $\omega$ is the fourier angular frequency
and $P_{\rm{trans}}$ is the transmitted laser power.
The averaged storage time of the cavity $t_c$
is given by $t_c = L \mathcal{F}/(\pi c)$,
where $L$ is the cavity round-trip length
and $\mathcal{F}$ is the finesse.
Simultaneous resonance of both carrier and sidebands beams
is also important for good sensitivity at low frequencies.

The signal-to-noise ratio improves with the measurement time $T^{1/2}$
as long as the axion oscillation is coherent for $T \lesssim \tau$, 
where $\tau$ is the coherent timescale of axion dark matter.
When the measurement time becomes longer than this coherence time $T \gtrsim \tau$,
the growth of the signal-to-noise ratio with the measurement time changes as $(T\tau)^{1/4}$.

\begin{figure}[h]
\begin{center}
\includegraphics[width=0.7\linewidth]{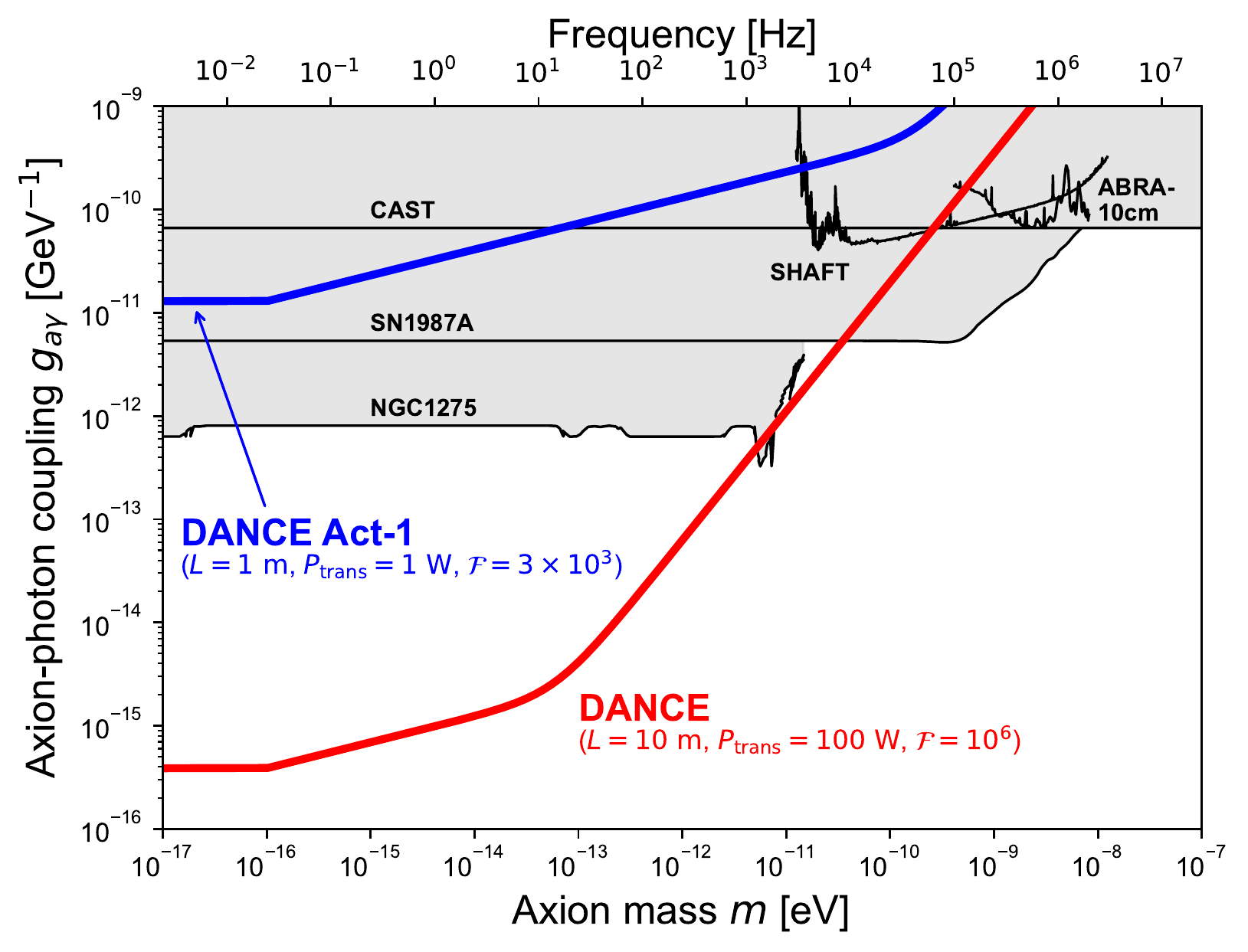}
\caption[]{The sensitivity curves 
for the axion-photon coupling constant $g_{a \gamma}$.
The blue (red) lines represent the designed shot noise limited sensitivity
of DANCE Act-1 (DANCE) with feasible (optimistic) parameters 
if we observe for a year.
The grey lines with shaded region are current bounds obtained from
CAST~\cite{CAST}, SHAFT~\cite{SHAFT}, and ABRACADABRA-10cm~\cite{ABRA-10cm} experiments,
and the astrophysical constraints from
the gamma-ray observations of SN1987A~\cite{SN1987A} 
and the X-ray observations of NGC1275 galaxy~\cite{NGC1275}.}
\label{fig:axionbounds}
\end{center}
\end{figure}

Assuming $L = \SI{10}{m}$, $\mathcal{F} = 10^6$,
and $P_{\rm{trans}} = \SI{100}{W}$,
we can reach
$g_{a \gamma} \simeq \SI{3e-16}{}\unit{GeV^{-1}}$
for $m_a < 10^{-16}\unit{eV}$
(see red line in Figure \ref{fig:axionbounds}).
Here, we set $\lambda = \SI{1064}{nm}$ 
and the integration time $T = 1 \unit{year}$.

\section{Prototype experiment DANCE Act-1}

\subsection{Experimental setups}

\begin{figure}[h]
\begin{minipage}{0.5\linewidth}
\centerline{\includegraphics[width=75mm]{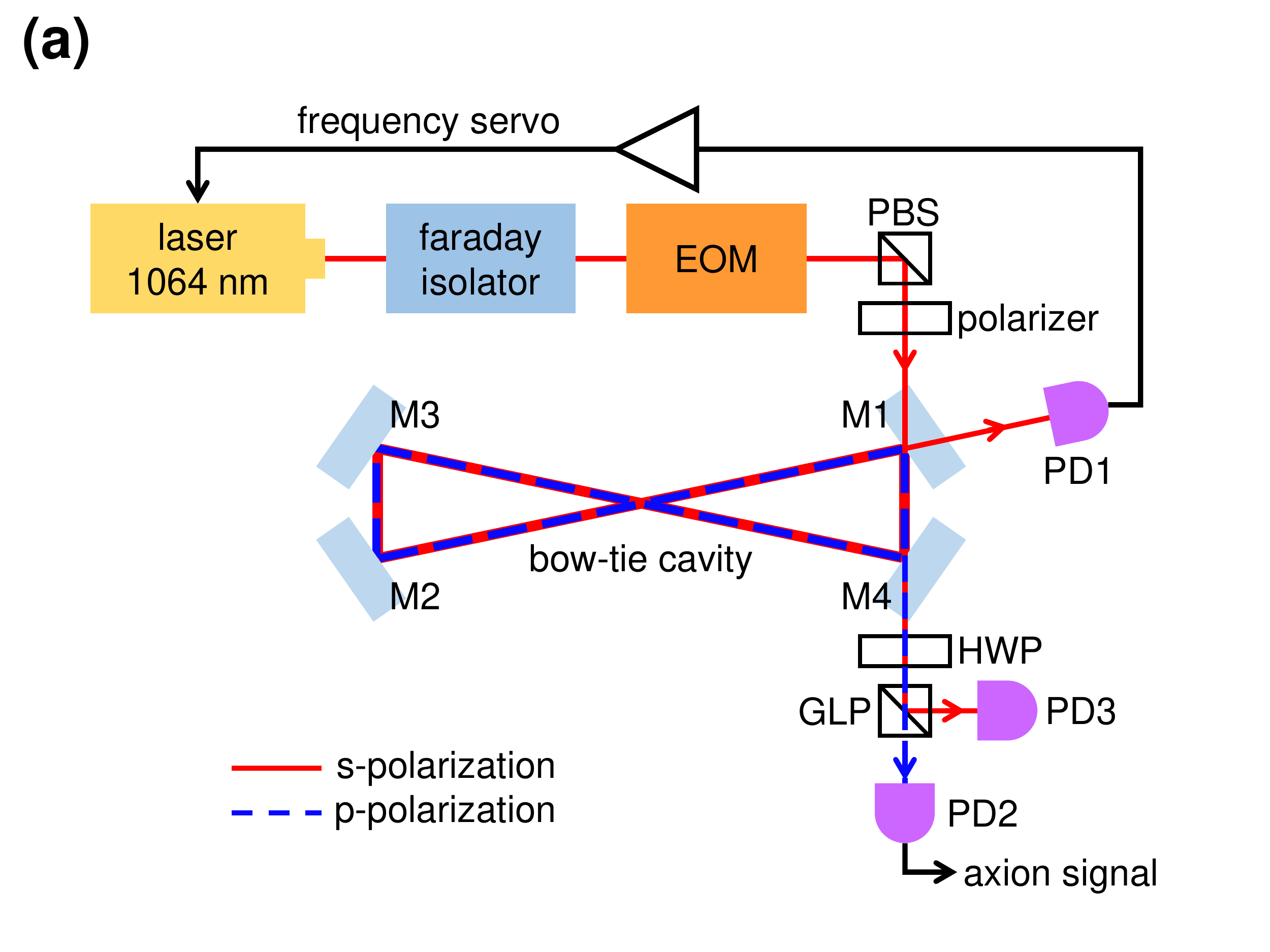}}
\end{minipage}
\hfill
\begin{minipage}{0.5\linewidth}
\centerline{\includegraphics[width=75mm]{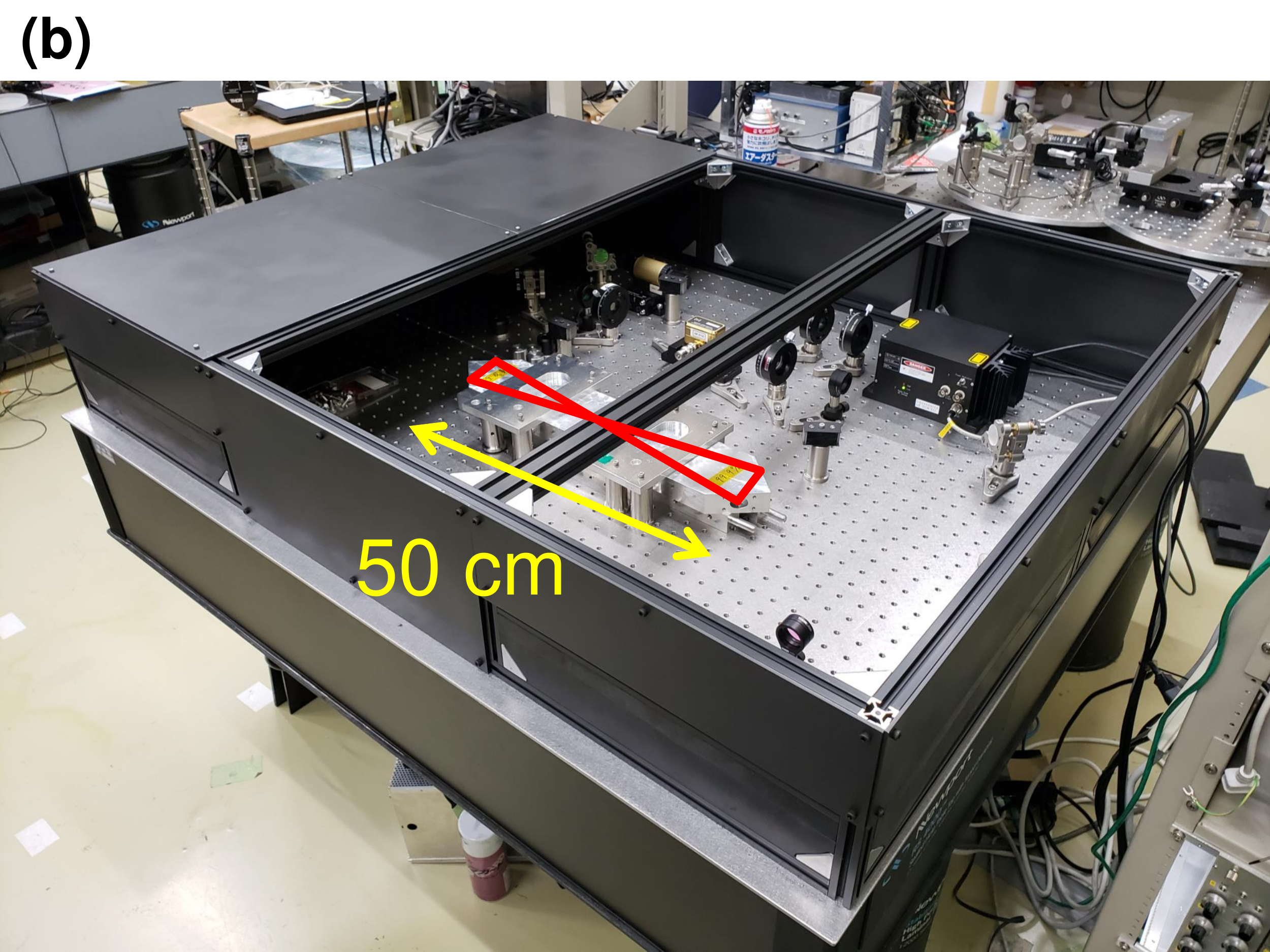}}
\end{minipage}
\caption[]{(a) The schematic of DANCE Act-1.
S-polarized (p-polarized) beam is drawn as red solid (blue dashed) lines.
EOM: electro-optic modulator.
PBS: polarizing beam splitter.
GLP: Glan laser polarizer.
PD: photodetector.
HWP: half-wave plate.
(b) A picture of DANCE Act-1.
}
\label{fig:setup}
\end{figure}

Since April 2019, we are continuing the prototype experiment DANCE Act-1~\cite{TAUP2019, Moriond_Oshima, Moriond_Fujimoto}.
Figure \ref{fig:setup} (a) shows the schematic of DANCE Act-1.
The s-polarized beam (the carrier in this work)
was fed into the bow-tie cavity
by putting a polarizing beam splitter (PBS) and a polarizer in front of the cavity.
The laser frequency was locked to the resonance of the bow-tie cavity
by the Pound-Drever-Hall technique.
Polarization of transmitted light was rotated with a half-wave plate (HWP)
to introduce some p-polarization (the sidebands in this work),
and then split
into p- and s-polarization with a Glan laser polarizer (GLP).
The amount of p- and s-polarization was recorded with photodetectors PD2 and PD3.

The length between mirrors M1 and M2 as well as M3 and M4 was $45 \unit{cm}$, 
and that between M2 and M3 as well as M4 and M1 was $4.7 \unit{cm}$.
Incident angles at all the four mirrors were $42 \unit{deg}$.
Mirrors M1-M4 were custom-made by Layertec.
All the four mirrors are concave mirrors with $\SI{1}{m}$ radius of curvature.
The power reflectivity for s-polarization of M1 and M4 was designed to be 99.90(2)\%
and that of M2 and M3 larger than $99.99\%$, 
which results in the designed finesse of \SI{3e3}{}.
We did not specify the reflectivity for p-polarization.

A photo of the experimental setup of DANCE Act-1
is shown in Figure \ref{fig:setup} (b).
The optical table is surrounded by aluminum plates
to stabilize the frequency control by reducing air turbulence
and to shield the optical setup from external light.
The bow-tie cavity is constructed from four mirrors
rigidly fixed on a spacer made of aluminum.

\subsection{Performance evaluation of the bow-tie cavity}

We evaluated the performance of the bow-tie cavity
by modulating the laser frequency 
and taking the cavity scan of transmitted light.
Results are summarized in Table \ref{tab:result}.
We injected laser power of around $\SI{200}{mW}$ into the cavity, which is lower than the designed power.
Transmitted laser power was lower than injected power
due to loss of light in the cavity.
Measured finesse for s-polarization
was consistent with the designed finesse.
Measured resonant frequencies for two polarizations were different
because p- and s-polarization obtained a non-zero phase shift
from mirror coating layers
when reflecting at oblique incident angles.
This resonant frequency difference degrades
the sensitivity of axion-photon coupling by two orders of magnitude
in the low mass region $m_a < 10^{-10}~\si{eV}$.
Note that smaller finesse for p-polarization is not an issue in this work
since the sensitivity in low mass region gets better with smaller finesse for p-polarization
when the resonant frequency difference between two polarizations is non-zero. 

\begin{table}[h]
\caption{\label{tab:result}Summary of the performance evaluation of the bow-tie cavity.}
\begin{center}
\begin{tabular}{lll}
\br
&Designed values&Measured values\\
\mr
Injected laser power&$\SI{1}{W}$&$\SI{242(12)}{mW}$\\
Transmitted laser power&$\SI{1}{W}$&$\SI{153(8)}{mW}$\\
Finesse for s-polarization (carrier)&$\SI{3.2(8)e3}{}$&$\SI{2.85(5)e3}{}$\\
Finesse for p-polarization (sidebands)&$-$&$195(3)$\\
Resonant frequency difference between polarizations&$\SI{0}{Hz}$&$\SI{2.52(2)}{MHz}$\\
\br
\end{tabular}
\end{center}
\end{table}

\subsection{First observation and sensitivity}

\begin{figure}[h]
\begin{center}
\includegraphics[width=0.63\linewidth]{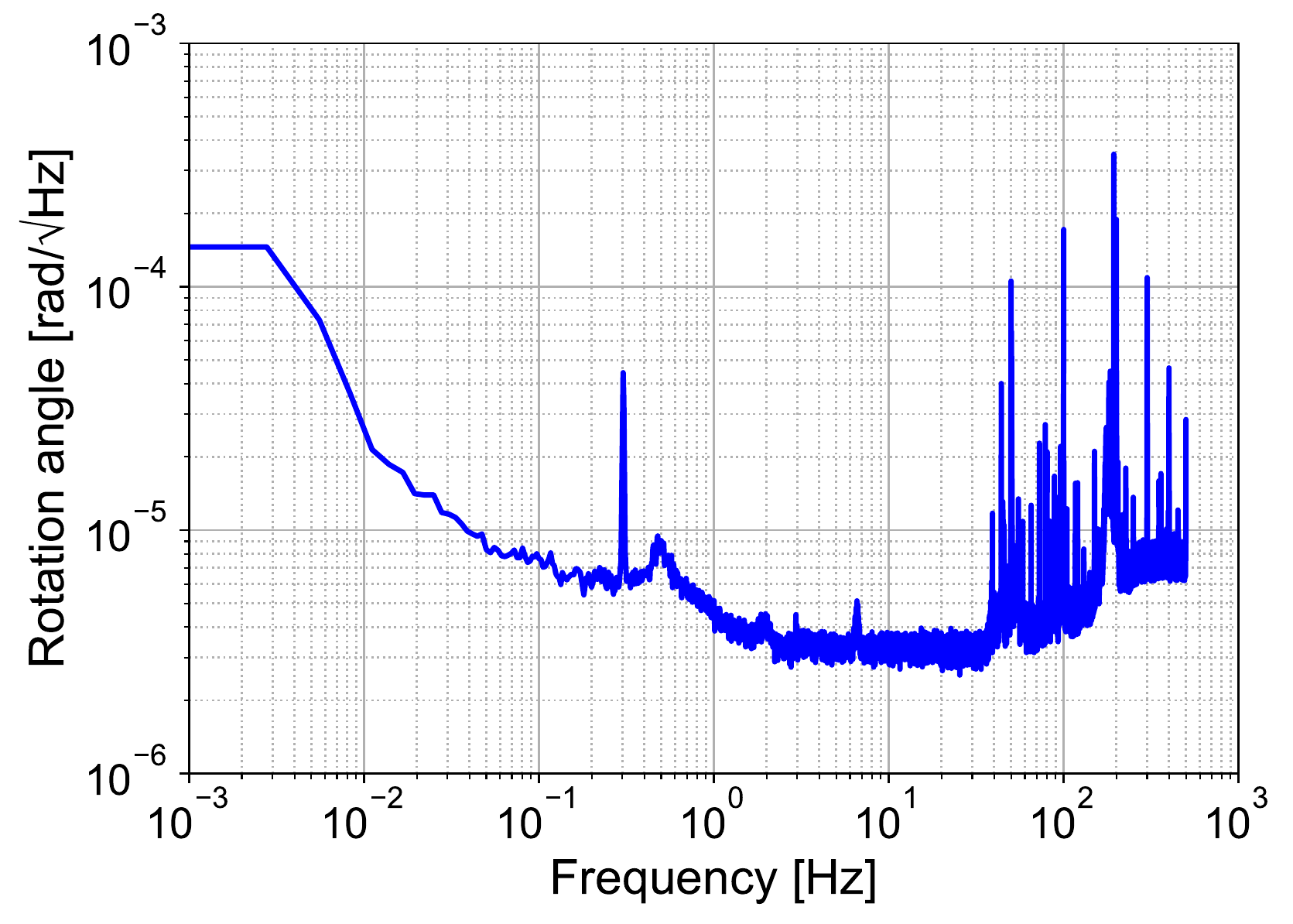}
\caption[]{The one-sided amplitude spectral density of the rotation angle of linear polarization.}
\label{fig:spectrum}
\end{center}
\end{figure}

The first data was taken for 12 days in May 2021.
The amount of p-polarization $P_{\rm{p}} (t)$ was observed with PD2,
and in this signal we can search for axions.
The amount of s-polarization was also measured with PD3 for calibration.
We calibrate the data to the rotation angle of linear polarization $\phi (t)$ by 
\begin{equation}
\phi (t) = \sqrt{\frac{P_{\rm{p}} (t)}{P_{\rm{tot}}}} - 2\theta,
\label{eq:rotangle}
\end{equation}
where $P_{\rm{tot}}$ is the averaged total amount of transmitted light
and $\theta$ is the fixed angle of the HWP.
The one-sided amplitude spectral density of the rotation angle of linear polarization 
is plotted in Figure \ref{fig:spectrum} by calculating with 10-hour data.
We reached $3 \times 10^{-6} \unit{rad/\sqrt{Hz}}$ at $\SI{10}{Hz}$.

Rotation angle of linear polarization
in $\SI{0.1}{Hz} - \SI{1}{Hz}$
correlated significantly
with injected laser power,
therefore the current sensitivity is believed to be probably limited
by laser intensity noise.
Whereas, 
rotation angle of linear polarization
in $\SI{30}{Hz} - \SI{5}{kHz}$
correlated significantly
with error signal for frequency servo,
therefore the current sensitivity seems to be limited
by mechanical vibration.

\section{Conclusion}

A new table-top experiment DANCE was proposed
to search for axion dark matter.
We aim to detect the rotational oscillation of linear polarization
caused by axion-photon coupling with the bow-tie cavity.
DANCE can improve the sensitivity
beyond the current bounds
of axion-photon coupling constant $g_{a \gamma}$
for axion mass $m_a < 10^{-10}~\si{eV}$
by several orders of magnitude.

A prototype experiment DANCE Act-1 is ongoing.
The assembly of the optics as well as the performance evaluation of the cavity has been completed.
Measured finesse for s-polarization (carrier) was consistent with the designed finesse,
while the resonant frequency difference between p- and s-polarizations was non-zero.
This leads to the degradation of the sensitivity by two orders of magnitude
in the low mass region $m_a < 10^{-10}~\si{eV}$.
We took the first data for 12 days in May 2021 and the data analysis is underway.
We have found candidate peaks for axions, 
but most of them turned out to be noise peaks.
We are working on further veto procedures and calibration to the coupling constant.

\section*{Acknowledgments}

We would like to thank Shigemi Otsuka and Togo Shimozawa
for manufacturing the mechanical parts,
and Ching Pin Ooi for editing this paper.
This work was supported by JSPS KAKENHI Grant Nos. 18H01224,
20H05850, 20H05854 and 20H05859, and JST PRESTO Grant No. JPMJPR200B.

\end{document}